\documentclass[12pt]{article}
\usepackage{epsf}
\usepackage{psfig}
\begin{document}

\input prepictex
\input pictexwd
\input postpictex

\begin{tabular}{l}
{\large {\bf Revisiting the Complexity of Finding Globally }} \\
{\large {\bf Minimum Energy Configurations in Atomic Clusters}} \\ \\
By G. W. Greenwood\\ \\
Dept. of Electrical \& Computer Engineering,
Portland State University, \\
Portland, OR 97207 USA
\end{tabular}

\vspace{0.2in}

\indent {\bf Published in \emph{Zeitschrift f\"{u}r
Physikalische Chemie} Vol. 211, 105-114, 1999}

\vspace{0.2in}

\begin{abstract}
It has previously been proven that finding the globally minimum 
energy configuration of an atomic cluster belongs in the class of 
NP-hard problems.  However, this proof is limited only to homonuclear 
clusters.  This paper presents a new proof which shows finding
minimum energy configurations for heteronuclear clusters is also 
NP-hard.
\end{abstract}

\vspace{0.2in}

\section{Introduction}

Atomic clusters are aggregates of atoms held together by the same forces 
that cause, for example, phase transition from vapor to liquid, formations 
of crystals, etc.  Cluster sizes range from as few as three atoms up to 
several hundred atoms.  The physical and chemical characteristics of a 
cluster often varies with its size.  In fact, even the addition of a 
single atom can result in an entirely different structure.  Only by 
successively adding more and more atoms will a crystal-like structure 
eventually be produced and some knowledge of the condensed phase 
attributes be determined \cite{mill}.

The study of atomic clusters has steadily been increasing over the past 
decade \cite{NBP}.  Of particular interest is the cluster conformation 
(structure) which has the lowest total internal energy $E$.  Knowledge 
of this minimum energy conformation provides valuable clues relating to 
the chemical and physical properties of the cluster.  Unfortunately, 
searching for the globally minimum energy state of a cluster has proven 
to be enormously difficult.  Indeed, Wille and Vennik \cite{wille85} 
showed that locating the globally minimum energy state of a cluster of 
identical atoms---the homonuclear case---belongs in the class of NP-hard 
problems.  This means there is little hope of exactly solving the problem in 
finite time for even moderate cluster sizes.

The purpose of this paper is two-fold.  First, it will be shown why 
existing homonuclear proofs, and work from other related problems, cannot
be used for the heteronuclear case where not all of the atoms are identical.  
Secondly, a proof will be presented which does show solving the heteronuclear 
problem is NP-hard.

\section{Preliminaries}

The problem of finding this globally minimum energy structure is equivalent 
to optimizing $E$ with respect to variations in all $3N - 6$ degrees of freedom
where $N$ is the cluster size.  One method of solving this optimization problem 
is to explore the potential energy surface (PES) composed of all possible cluster 
conformations.  Each point on this surface represents a unique spatial 
arrangement of the constituent atoms.  This multidimensional surface is 
characterized by numerous local minima, each indicating an energetically 
stable structure.  If it were possible to enumerate all of these minima---and 
the saddles that link them---we could, in principle, describe the dynamics of 
chemical reactions governed by that surface.  Unfortunately, enumerating all 
minima is extremely difficult because of their large number.  Berry \cite{berry93} 
indicates the number of geometrically distinct minima tends to grow exponentially 
with $N$.  Moreover, the number of permutational isomers grows 
factorially with $N$.

A number of general methods have been proposed for finding global minima on a PES, 
in particular, and on hypersurfaces in general.  For example, Monte Carlo methods 
\cite{pang94}, eigenvector following \cite{banerjee85}, evolution computation 
techniques \cite{gree98}-\cite{zeiri95}, lattice optimization/relaxation techniques 
\cite{north87}, and PES deformation techniques \cite{kost91,wales97} have all been 
used with varying degrees of success.  After formally defining the clustering problem, 
we will discuss some of these techniques in greater detail.

The potential energy function for a cluster of $N$ atoms is given by
\begin{equation}
V(r^N) \; = \; {\frac {1}{2}} \sum_{\footnotesize 
\begin{array}{c}i,j\\i\ne j\end{array}}^N
v(r_i - r_j)
\label{eqn1}
\end{equation}
where $r^N=(r_1,r_2,\ldots,r_N)$,
$r_i$ is the position vector of the $i$th atom and $v(r_i-r_j)$ is a function 
representing the pairwise interaction between atoms.  Lennard-Jones or Morse 
functions are frequently used for these functions.  Our optimization problem 
of interest is the Discrete Cluster Problem (DCP) which is formally defined as 
follows:\footnote{The definition given covers both homonuclear and heteronuclear
systems.}

\vspace{0.1in}

\noindent {\bf DISCRETE CLUSTER PROBLEM}

\noindent INSTANCE: Given finite number of points in real space, a distance 
$d(i,j)\in Z^+$ between two points $i,j$, an integer $N\in Z^+$ and a known 
potential energy function defined by (\ref{eqn1}). 

\noindent QUESTION:  Is there a way to assign $N$ atoms to $N$ unique points 
so as to minimize the sum of their pairwise interactions?

\vspace{0.2in}

The potential energy of a cluster actually equals the sum of $N$-body
interactions.  Restricting this sum to only two-body interactions provides
only a \emph{qualitative} approximation.  Three-body interactions can be 
important, but are usually ignored in first order approximations \cite{klein90}.
Although many-body interactions are needed for \emph{quantitative} modeling,
little is actually known about higher order terms.  Many-body potential energy
functions do exist \cite{erkoc97}, though in practice, only two-body terms are
used for the sake of computational speed.  Consequently, the discussion here will
likewise be restricted to the two-body case.

\begin{figure}[htbp]
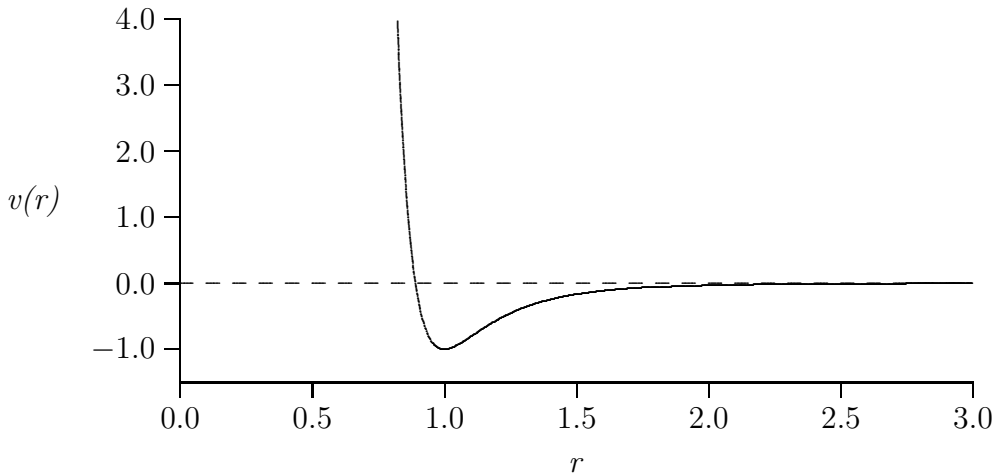

\[\hspace*{-0.7in}\beginpicture
    \setcoordinatesystem units <100pt,25pt>
    \setplotarea x from 0 to 3, y from -1.5 to 4.0 
    \axis bottom label {\emph{r}} 
     ticks numbered from 0 to 3 by 0.5 /
    \axis left label {\emph{v(r)}}
     ticks numbered from -1.0 to 4.0 by 1.0 /

\setsolid\plot 
0.822499 3.972376
0.824999 3.715847
0.827499 3.471287
0.829999 3.238150
0.832499 3.015924
0.834999 2.804121
0.837499 2.602270
0.839999 2.409929
0.842499 2.226669
0.844999 2.052081
0.847499 1.885780
0.849999 1.727393
0.852499 1.576564
0.854999 1.432963
0.857499 1.296266
0.859999 1.166162
0.862499 1.042359
0.864999 0.924580
0.867499 0.812548
0.869999 0.706015
0.872499 0.604733
0.874999 0.508470
0.877499 0.417003
0.879999 0.330119
0.882499 0.247614
0.884999 0.169291
0.887499 0.094966
0.889999 0.024464
0.892499 -0.042389
0.894999 -0.105753
0.897499 -0.165784
0.899999 -0.222631
0.902499 -0.276434
0.904999 -0.327329
0.907499 -0.375448
0.909999 -0.420913
0.912499 -0.463842
0.914999 -0.504351
0.917499 -0.542545
0.919999 -0.578530
0.922499 -0.612404
0.924999 -0.644263
0.927499 -0.674196
0.929999 -0.702292
0.932499 -0.728632
0.934999 -0.753297
0.937499 -0.776362
0.939999 -0.797900
0.942499 -0.817980
0.944999 -0.836669
0.947499 -0.854031
0.949999 -0.870126
0.952499 -0.885013
0.954999 -0.898748
0.957499 -0.911383
0.959999 -0.922971
0.962499 -0.933560
0.964999 -0.943196
0.967499 -0.951925
0.969999 -0.959789
0.972499 -0.966829
0.974999 -0.973085
0.977499 -0.978594
0.979999 -0.983391
0.982499 -0.987512
0.984999 -0.990989
0.987499 -0.993854
0.989999 -0.996136
0.992499 -0.997865
0.994999 -0.999068
0.997499 -0.999771
0.999999 -1.000000
1.002499 -0.999779
1.004999 -0.999131
1.007499 -0.998078
1.009999 -0.996642
1.012499 -0.994842
1.015000 -0.992697
1.017500 -0.990228
1.020000 -0.987450
1.022500 -0.984381
1.025000 -0.981038
1.027500 -0.977436
1.030000 -0.973589
1.032500 -0.969512
1.035000 -0.965218
1.037500 -0.960721
1.040000 -0.956032
1.042500 -0.951164
1.045000 -0.946127
1.047500 -0.940933
1.050000 -0.935593
1.052500 -0.930115
1.055000 -0.924509
1.057500 -0.918785
1.060001 -0.912950
1.062501 -0.907014
1.065001 -0.900984
1.067501 -0.894867
1.070001 -0.888671
1.072501 -0.882402
1.075001 -0.876067
1.077501 -0.869672
1.080001 -0.863223
1.082501 -0.856726
1.085001 -0.850186
1.087501 -0.843608
1.090001 -0.836997
1.092501 -0.830357
1.095001 -0.823694
1.097501 -0.817012
1.100001 -0.810313
1.102502 -0.803603
1.105002 -0.796884
1.107502 -0.790161
1.110002 -0.783436
1.112502 -0.776713
1.115002 -0.769994
1.117502 -0.763283
1.120002 -0.756582
1.122502 -0.749893
1.125002 -0.743219
1.127502 -0.736563
1.130002 -0.729926
1.132502 -0.723309
1.135002 -0.716717
1.137502 -0.710149
1.140002 -0.703608
1.142502 -0.697095
1.145002 -0.690612
1.147503 -0.684160
1.150003 -0.677741
1.152503 -0.671356
1.155003 -0.665006
1.157503 -0.658692
1.160003 -0.652415
1.162503 -0.646176
1.165003 -0.639975
1.167503 -0.633815
1.170003 -0.627696
1.172503 -0.621617
1.175003 -0.615581
1.177503 -0.609587
1.180003 -0.603636
1.182503 -0.597728
1.185003 -0.591865
1.187503 -0.586046
1.190004 -0.580272
1.192504 -0.574544
1.195004 -0.568860
1.197504 -0.563223
1.200004 -0.557631
1.202504 -0.552086
1.205004 -0.546586
1.207504 -0.541134
1.210004 -0.535727
1.212504 -0.530368
1.215004 -0.525055
1.217504 -0.519789
1.220004 -0.514570
1.222504 -0.509397
1.225004 -0.504271
1.227504 -0.499192
1.230004 -0.494159
1.232504 -0.489173
1.235005 -0.484233
1.237505 -0.479340
1.240005 -0.474492
1.242505 -0.469690
1.245005 -0.464935
1.247505 -0.460224
1.250005 -0.455559
1.252505 -0.450940
1.255005 -0.446365
1.257505 -0.441835
1.260005 -0.437350
1.262505 -0.432909
1.265005 -0.428512
1.267505 -0.424159
1.270005 -0.419849
1.272505 -0.415582
1.275005 -0.411359
1.277506 -0.407178
1.280006 -0.403039
1.282506 -0.398943
1.285006 -0.394888
1.287506 -0.390875
1.290006 -0.386903
1.292506 -0.382972
1.295006 -0.379082
1.297506 -0.375232
1.300006 -0.371421
1.302506 -0.367651
1.305006 -0.363919
1.307506 -0.360227
1.310006 -0.356573
1.312506 -0.352957
1.315006 -0.349379
1.317506 -0.345840
1.320006 -0.342337
1.322507 -0.338871
1.325007 -0.335442
1.327507 -0.332049
1.330007 -0.328692
1.332507 -0.325371
1.335007 -0.322085
1.337507 -0.318834
1.340007 -0.315618
1.342507 -0.312436
1.345007 -0.309288
1.347507 -0.306173
1.350007 -0.303092
1.352507 -0.300044
1.355007 -0.297029
1.357507 -0.294045
1.360007 -0.291094
1.362507 -0.288175
1.365008 -0.285287
1.367508 -0.282430
1.370008 -0.279604
1.372508 -0.276808
1.375008 -0.274042
1.377508 -0.271306
1.380008 -0.268600
1.382508 -0.265923
1.385008 -0.263274
1.387508 -0.260655
1.390008 -0.258063
1.392508 -0.255500
1.395008 -0.252964
1.397508 -0.250455
1.400008 -0.247974
1.402508 -0.245519
1.405008 -0.243091
1.407508 -0.240689
1.410009 -0.238313
1.412509 -0.235963
1.415009 -0.233638
1.417509 -0.231339
1.420009 -0.229064
1.422509 -0.226813
1.425009 -0.224587
1.427509 -0.222385
1.430009 -0.220207
1.432509 -0.218052
1.435009 -0.215921
1.437509 -0.213812
1.440009 -0.211727
1.442509 -0.209663
1.445009 -0.207622
1.447509 -0.205603
1.450009 -0.203606
1.452510 -0.201630
1.455010 -0.199675
1.457510 -0.197742
1.460010 -0.195829
1.462510 -0.193937
1.465010 -0.192065
1.467510 -0.190213
1.470010 -0.188381
1.472510 -0.186569
1.475010 -0.184776
1.477510 -0.183002
1.480010 -0.181248
1.482510 -0.179512
1.485010 -0.177794
1.487510 -0.176095
1.490010 -0.174415
1.492510 -0.172752
1.495010 -0.171107
1.497511 -0.169479
1.500011 -0.167869
1.502511 -0.166276
1.505011 -0.164700
1.507511 -0.163140
1.510011 -0.161597
1.512511 -0.160071
1.515011 -0.158561
1.517511 -0.157066
1.520011 -0.155588
1.522511 -0.154125
1.525011 -0.152678
1.527511 -0.151246
1.530011 -0.149829
1.532511 -0.148427
1.535011 -0.147040
1.537511 -0.145668
1.540012 -0.144310
1.542512 -0.142966
1.545012 -0.141636
1.547512 -0.140320
1.550012 -0.139018
1.552512 -0.137730
1.555012 -0.136455
1.557512 -0.135194
1.560012 -0.133946
1.562512 -0.132711
1.565012 -0.131488
1.567512 -0.130279
1.570012 -0.129082
1.572512 -0.127897
1.575012 -0.126725
1.577512 -0.125565
1.580012 -0.124417
1.582512 -0.123281
1.585013 -0.122157
1.587513 -0.121044
1.590013 -0.119943
1.592513 -0.118853
1.595013 -0.117775
1.597513 -0.116708
1.600013 -0.115651
1.602513 -0.114606
1.605013 -0.113571
1.607513 -0.112547
1.610013 -0.111533
1.612513 -0.110530
1.615013 -0.109537
1.617513 -0.108554
1.620013 -0.107581
1.622513 -0.106618
1.625013 -0.105665
1.627514 -0.104722
1.630014 -0.103788
1.632514 -0.102864
1.635014 -0.101949
1.637514 -0.101044
1.640014 -0.100147
1.642514 -0.099260
1.645014 -0.098382
1.647514 -0.097512
1.650014 -0.096651
1.652514 -0.095799
1.655014 -0.094956
1.657514 -0.094121
1.660014 -0.093294
1.662514 -0.092476
1.665014 -0.091666
1.667514 -0.090864
1.670015 -0.090070
1.672515 -0.089284
1.675015 -0.088506
1.677515 -0.087736
1.680015 -0.086973
1.682515 -0.086218
1.685015 -0.085470
1.687515 -0.084730
1.690015 -0.083997
1.692515 -0.083272
1.695015 -0.082553
1.697515 -0.081842
1.700015 -0.081138
1.702515 -0.080440
1.705015 -0.079750
1.707515 -0.079066
1.710015 -0.078389
1.712515 -0.077719
1.715016 -0.077055
1.717516 -0.076398
1.720016 -0.075747
1.722516 -0.075103
1.725016 -0.074465
1.727516 -0.073833
1.730016 -0.073207
1.732516 -0.072588
1.735016 -0.071974
1.737516 -0.071366
1.740016 -0.070764
1.742516 -0.070168
1.745016 -0.069578
1.747516 -0.068994
1.750016 -0.068415
1.752516 -0.067842
1.755016 -0.067274
1.757517 -0.066712
1.760017 -0.066155
1.762517 -0.065603
1.765017 -0.065057
1.767517 -0.064516
1.770017 -0.063980
1.772517 -0.063450
1.775017 -0.062924
1.777517 -0.062403
1.780017 -0.061888
1.782517 -0.061377
1.785017 -0.060871
1.787517 -0.060370
1.790017 -0.059874
1.792517 -0.059382
1.795017 -0.058895
1.797517 -0.058412
1.800017 -0.057935
1.802518 -0.057461
1.805018 -0.056992
1.807518 -0.056528
1.810018 -0.056068
1.812518 -0.055612
1.815018 -0.055160
1.817518 -0.054713
1.820018 -0.054270
1.822518 -0.053831
1.825018 -0.053396
1.827518 -0.052965
1.830018 -0.052538
1.832518 -0.052116
1.835018 -0.051697
1.837518 -0.051282
1.840018 -0.050870
1.842518 -0.050463
1.845019 -0.050059
1.847519 -0.049659
1.850019 -0.049263
1.852519 -0.048871
1.855019 -0.048482
1.857519 -0.048096
1.860019 -0.047714
1.862519 -0.047336
1.865019 -0.046961
1.867519 -0.046590
1.870019 -0.046222
1.872519 -0.045857
1.875019 -0.045496
1.877519 -0.045138
1.880019 -0.044783
1.882519 -0.044431
1.885019 -0.044083
1.887519 -0.043737
1.890020 -0.043395
1.892520 -0.043056
1.895020 -0.042720
1.897520 -0.042387
1.900020 -0.042057
1.902520 -0.041730
1.905020 -0.041406
1.907520 -0.041085
1.910020 -0.040767
1.912520 -0.040451
1.915020 -0.040139
1.917520 -0.039829
1.920020 -0.039522
1.922520 -0.039218
1.925020 -0.038916
1.927520 -0.038617
1.930020 -0.038321
1.932521 -0.038027
1.935021 -0.037736
1.937521 -0.037448
1.940021 -0.037162
1.942521 -0.036879
1.945021 -0.036598
1.947521 -0.036319
1.950021 -0.036043
1.952521 -0.035770
1.955021 -0.035499
1.957521 -0.035230
1.960021 -0.034964
1.962521 -0.034700
1.965021 -0.034438
1.967521 -0.034179
1.970021 -0.033921
1.972521 -0.033666
1.975021 -0.033414
1.977522 -0.033163
1.980022 -0.032915
1.982522 -0.032669
1.985022 -0.032425
1.987522 -0.032183
1.990022 -0.031943
1.992522 -0.031705
1.995022 -0.031469
1.997522 -0.031236
2.000022 -0.031004
2.002522 -0.030774
2.005022 -0.030546
2.007522 -0.030321
2.010022 -0.030097
2.012522 -0.029875
2.015022 -0.029655
2.017522 -0.029437
2.020022 -0.029220
2.022522 -0.029006
2.025023 -0.028793
2.027523 -0.028582
2.030023 -0.028373
2.032523 -0.028166
2.035023 -0.027961
2.037523 -0.027757
2.040023 -0.027555
2.042523 -0.027354
2.045023 -0.027156
2.047523 -0.026959
2.050023 -0.026763
2.052523 -0.026570
2.055023 -0.026378
2.057523 -0.026187
2.060023 -0.025998
2.062523 -0.025811
2.065023 -0.025625
2.067523 -0.025441
2.070024 -0.025259
2.072524 -0.025078
2.075024 -0.024898
2.077524 -0.024720
2.080024 -0.024543
2.082524 -0.024368
2.085024 -0.024194
2.087524 -0.024022
2.090024 -0.023851
2.092524 -0.023682
2.095024 -0.023514
2.097524 -0.023347
2.100024 -0.023182
2.102524 -0.023018
2.105024 -0.022855
2.107524 -0.022694
2.110024 -0.022534
2.112525 -0.022375
2.115025 -0.022218
2.117525 -0.022062
2.120025 -0.021907
2.122525 -0.021754
2.125025 -0.021601
2.127525 -0.021450
2.130025 -0.021301
2.132525 -0.021152
2.135025 -0.021005
2.137525 -0.020858
2.140025 -0.020713
2.142525 -0.020569
2.145025 -0.020427
2.147525 -0.020285
2.150025 -0.020145
2.152525 -0.020006
2.155025 -0.019867
2.157526 -0.019730
2.160026 -0.019594
2.162526 -0.019460
2.165026 -0.019326
2.167526 -0.019193
2.170026 -0.019061
2.172526 -0.018931
2.175026 -0.018801
2.177526 -0.018673
2.180026 -0.018545
2.182526 -0.018419
2.185026 -0.018293
2.187526 -0.018169
2.190026 -0.018045
2.192526 -0.017923
2.195026 -0.017801
2.197526 -0.017680
2.200027 -0.017561
2.202527 -0.017442
2.205027 -0.017324
2.207527 -0.017207
2.210027 -0.017091
2.212527 -0.016976
2.215027 -0.016862
2.217527 -0.016749
2.220027 -0.016637
2.222527 -0.016525
2.225027 -0.016414
2.227527 -0.016305
2.230027 -0.016196
2.232527 -0.016088
2.235027 -0.015980
2.237527 -0.015874
2.240027 -0.015768
2.242527 -0.015664
2.245028 -0.015560
2.247528 -0.015457
2.250028 -0.015354
2.252528 -0.015253
2.255028 -0.015152
2.257528 -0.015052
2.260028 -0.014953
2.262528 -0.014854
2.265028 -0.014756
2.267528 -0.014659
2.270028 -0.014563
2.272528 -0.014468
2.275028 -0.014373
2.277528 -0.014279
2.280028 -0.014185
2.282528 -0.014093
2.285028 -0.014001
2.287529 -0.013910
2.290029 -0.013819
2.292529 -0.013729
2.295029 -0.013640
2.297529 -0.013551
2.300029 -0.013464
2.302529 -0.013376
2.305029 -0.013290
2.307529 -0.013204
2.310029 -0.013119
2.312529 -0.013034
2.315029 -0.012950
2.317529 -0.012867
2.320029 -0.012784
2.322529 -0.012702
2.325029 -0.012621
2.327529 -0.012540
2.330029 -0.012460
2.332530 -0.012380
2.335030 -0.012301
2.337530 -0.012222
2.340030 -0.012144
2.342530 -0.012067
2.345030 -0.011990
2.347530 -0.011914
2.350030 -0.011839
2.352530 -0.011763
2.355030 -0.011689
2.357530 -0.011615
2.360030 -0.011542
2.362530 -0.011469
2.365030 -0.011396
2.367530 -0.011325
2.370030 -0.011253
2.372530 -0.011183
2.375031 -0.011112
2.377531 -0.011043
2.380031 -0.010973
2.382531 -0.010905
2.385031 -0.010836
2.387531 -0.010769
2.390031 -0.010701
2.392531 -0.010635
2.395031 -0.010568
2.397531 -0.010503
2.400031 -0.010437
2.402531 -0.010373
2.405031 -0.010308
2.407531 -0.010244
2.410031 -0.010181
2.412531 -0.010118
2.415031 -0.010055
2.417531 -0.009993
2.420032 -0.009932
2.422532 -0.009870
2.425032 -0.009810
2.427532 -0.009749
2.430032 -0.009690
2.432532 -0.009630
2.435032 -0.009571
2.437532 -0.009512
2.440032 -0.009454
2.442532 -0.009396
2.445032 -0.009339
2.447532 -0.009282
2.450032 -0.009226
2.452532 -0.009169
2.455032 -0.009114
2.457532 -0.009058
2.460032 -0.009003
2.462533 -0.008949
2.465033 -0.008895
2.467533 -0.008841
2.470033 -0.008787
2.472533 -0.008734
2.475033 -0.008682
2.477533 -0.008629
2.480033 -0.008577
2.482533 -0.008526
2.485033 -0.008474
2.487533 -0.008424
2.490033 -0.008373
2.492533 -0.008323
2.495033 -0.008273
2.497533 -0.008224
2.500033 -0.008175
2.502533 -0.008126
2.505033 -0.008077
2.507534 -0.008029
2.510034 -0.007981
2.512534 -0.007934
2.515034 -0.007887
2.517534 -0.007840
2.520034 -0.007794
2.522534 -0.007748
2.525034 -0.007702
2.527534 -0.007656
2.530034 -0.007611
2.532534 -0.007566
2.535034 -0.007522
2.537534 -0.007477
2.540034 -0.007433
2.542534 -0.007390
2.545034 -0.007346
2.547534 -0.007303
2.550035 -0.007260
2.552535 -0.007218
2.555035 -0.007176
2.557535 -0.007134
2.560035 -0.007092
2.562535 -0.007051
2.565035 -0.007010
2.567535 -0.006969
2.570035 -0.006929
2.572535 -0.006888
2.575035 -0.006848
2.577535 -0.006809
2.580035 -0.006769
2.582535 -0.006730
2.585035 -0.006691
2.587535 -0.006653
2.590035 -0.006614
2.592535 -0.006576
2.595036 -0.006538
2.597536 -0.006501
2.600036 -0.006463
2.602536 -0.006426
2.605036 -0.006389
2.607536 -0.006353
2.610036 -0.006316
2.612536 -0.006280
2.615036 -0.006244
2.617536 -0.006209
2.620036 -0.006173
2.622536 -0.006138
2.625036 -0.006103
2.627536 -0.006068
2.630036 -0.006034
2.632536 -0.006000
2.635036 -0.005966
2.637537 -0.005932
2.640037 -0.005898
2.642537 -0.005865
2.645037 -0.005832
2.647537 -0.005799
2.650037 -0.005766
2.652537 -0.005734
2.655037 -0.005701
2.657537 -0.005669
2.660037 -0.005638
2.662537 -0.005606
2.665037 -0.005574
2.667537 -0.005543
2.670037 -0.005512
2.672537 -0.005481
2.675037 -0.005451
2.677537 -0.005420
2.680037 -0.005390
2.682538 -0.005360
2.685038 -0.005330
2.687538 -0.005301
2.690038 -0.005271
2.692538 -0.005242
2.695038 -0.005213
2.697538 -0.005184
2.700038 -0.005155
2.702538 -0.005127
2.705038 -0.005098
2.707538 -0.005070
2.710038 -0.005042
2.712538 -0.005015
2.715038 -0.004987
2.717538 -0.004959
2.720038 -0.004932
2.722538 -0.004905
2.725039 -0.004878
2.727539 -0.004852
2.730039 -0.004825
2.732539 -0.004799
2.735039 -0.004772
2.737539 -0.004746
2.740039 -0.004720
2.742539 -0.004695
2.745039 -0.004669
2.747539 -0.004644
2.750039 -0.004618
2.752539 -0.004593
2.755039 -0.004568
2.757539 -0.004544
2.760039 -0.004519
2.762539 -0.004495
2.765039 -0.004470
2.767540 -0.004446
2.770040 -0.004422
2.772540 -0.004398
2.775040 -0.004375
2.777540 -0.004351
2.780040 -0.004328
2.782540 -0.004304
2.785040 -0.004281
2.787540 -0.004258
2.790040 -0.004236
2.792540 -0.004213
2.795040 -0.004190
2.797540 -0.004168
2.800040 -0.004146
2.802540 -0.004124
2.805040 -0.004102
2.807540 -0.004080
2.810040 -0.004058
2.812541 -0.004036
2.815041 -0.004015
2.817541 -0.003994
2.820041 -0.003973
2.822541 -0.003951
2.825041 -0.003931
2.827541 -0.003910
2.830041 -0.003889
2.832541 -0.003869
2.835041 -0.003848
2.837541 -0.003828
2.840041 -0.003808
2.842541 -0.003788
2.845041 -0.003768
2.847541 -0.003748
2.850041 -0.003728
2.852541 -0.003709
2.855042 -0.003689
2.857542 -0.003670
2.860042 -0.003651
2.862542 -0.003632
2.865042 -0.003613
2.867542 -0.003594
2.870042 -0.003575
2.872542 -0.003557
2.875042 -0.003538
2.877542 -0.003520
2.880042 -0.003502
2.882542 -0.003483
2.885042 -0.003465
2.887542 -0.003447
2.890042 -0.003430
2.892542 -0.003412
2.895042 -0.003394
2.897542 -0.003377
2.900043 -0.003359
2.902543 -0.003342
2.905043 -0.003325
2.907543 -0.003308
2.910043 -0.003291
2.912543 -0.003274
2.915043 -0.003257
2.917543 -0.003240
2.920043 -0.003224
2.922543 -0.003207
2.925043 -0.003191
2.927543 -0.003174
2.930043 -0.003158
2.932543 -0.003142
2.935043 -0.003126
2.937543 -0.003110
2.940043 -0.003094
2.942544 -0.003079
2.945044 -0.003063
2.947544 -0.003047
2.950044 -0.003032
2.952544 -0.003017
2.955044 -0.003001
2.957544 -0.002986
2.960044 -0.002971
2.962544 -0.002956
2.965044 -0.002941
2.967544 -0.002926
2.970044 -0.002912
2.972544 -0.002897
2.975044 -0.002882
2.977544 -0.002868
2.980044 -0.002854
2.982544 -0.002839
2.985044 -0.002825
2.987545 -0.002811
2.990045 -0.002797
2.992545 -0.002783
2.995045 -0.002769
2.997545 -0.002755
/
\setdashes\plot
0.0 0.0
3.0 0.0
/

\endpicture\]
\caption{
The scaled Lennard-Jones potential function $v(r)$.  The 
scaled interatomic distance is denoted by $r$.}
\label{lj}
\end{figure}

It may appear that the large amount of work done with hard-sphere packing 
problems will be helpful in solving instances of DCP.  Unfortunately, such
is not the case because the objective of the two problems are quite different.  
Hard-sphere packings try to place $N$ spheres in 
Euclidean space so that all can, without overlap, fit within as small a volume 
as possible \cite{stewart92}.   DCP deals with soft, compliant spheres which
interact via pairwise interaction functions.  Figure \ref{lj} shows a 
Lennard-Jones function, which is typical.  Therein lies the major difference
between hard-sphere and soft-sphere systems: the former has no preferred 
distance between the spheres while the latter does.  Put another way,
a hard-sphere packing algorithm attempts to minimize the interatomic distance $r$.
Yet, a comparison with Figure \ref{lj} clearly shows this does not yield the 
lowest energy state for an atomic pair.  Hard-sphere packing studies can thus 
be expected to provide little help.

An algorithm that searches for solutions to the homonuclear 
version of DCP was recently proposed by Hendrickson \cite{hen95}.
The general approach exploits structural information to decompose 
the global optimization problem into a set of smaller optimization 
problems.  More specifically, $N$ objects in Euclidean space
are represented as an undirected graph where the edge weights 
reflect the interobject distance.  A subgraph is identified, the relative
positions of the vertices are optimized, and the subgraph is
then treated as a rigid body.  These rigid bodies 
are ultimately combined to determine the overall structure.  
Such an approach won't work for the heteronuclear case, but this discussion
will be differed until Section \ref{3}.

Northby \cite{north87} performed a lattice based search that takes an initial
conformation and allows it to relax to an energy minimum.  Essentially, a random 
search is conducted to compile a list of isomers.  After pruning any geometrically 
equivalent isomers, the remaining conformations are relaxed under a Lennard-Jones 
pairwise interaction function.  While this technique will work with heteronuclear
clusters, it does depend upon a gradient search which makes it susceptible to 
stopping at local optima.

Kostrowicki et al.~\cite{kost91},  presented a technique that
``deforms'' the PES in such a way that the number of local minima is dramatically
reduced.  This deformation requires solving a partial differential equation called 
a diffusion equation.  All atoms are assumed to interact by a Lennard-Jones type
of function which is approximated by a linear combination of Gaussian functions
to permit expressing the answer analytically.  Solving these diffusion equations
is not necessarily easy as there are some pathological conditions.  For example, 
numerical problems can occur when the atoms are  separated by large distances
and so interact only weakly.  This technique works best if one can suitably 
modify the boundary conditions for the diffusion equation, but care must be taken 
to ensure the minimum energy state is not removed from the deformed PES.

Wales and Doye \cite{wales97} described another PES transformation technique
called ``basin hopping''.  Here the PES is converted into a set of basins of
attraction for the local minima.  This approach does has the advantage of not
altering the energies of the minima.  It has been used to find many of the
local minima for Lennard-Jones cluster sizes up to $N=110$,  The authors 
suggest some improvements that could be made to improve the efficiency of
their approach.

\emph{It is important to note that none of these techniques will, with certainty,
guarantee a successful search for the globally minimum energy configuration}.  In 
the next section we will see why solving instances of DCP has proven to be so difficult.

\section{Complexity in Homonuclear Clusters}  \label{3}

An instance of DCP can be solved by exploring the PES associated the cluster, and 
any algorithm that does this can be used with both homonuclear and heteronuclear systems. 
A number of researchers indicate the number of local optima in a PES grows 
exponentially with $N$ \cite{berry93,hoare79,ball96}.  So, exactly how much effort is 
required to explore an exponentially large hypersurface? 
Some insight can be found from the theory of NP-completeness \cite{garey79}.

Suppose each time step a new point on the hypersurface is visited.  Repeating
this process until all points are visited---a procedure guaranteed to find the 
minimum energy state---will take an exponential number of time steps.  Now consider
an arbitrary NP-complete problem ${\cal P}$.  
If we found an algorithm that 
could solve ${\cal P}$ in polynomial time, then this algorithm could
solve every NP-complete problem in polynomial time.  Unfortunately, no
algorithm that solves an NP-complete problem has ever been found which
executes in less than an exponential number of steps.  This suggests the
effort required to conduct a blind search of the hypersurface is similar 
to the effort required to solve an NP-complete problem.  NP-complete
problems are computationally intractable, which gives some
idea of the difficulty one faces in solving DCP.  (NOTE: This
does \emph{not} prove DCP is NP-complete.  It merely shows 
the ramifications of having so many local optima.)

NP-complete problems are defined as decision problems, that is,
problems answered by either `yes' or `no'.  NP-hard problems ask for 
the optimal solution to an NP-complete problem.  And, they have
at least the same level of difficulty to solve as does the
corresponding NP-complete problem.  One particular NP-hard problem 
plays a pivotal role in the complexity analysis of DCP.  It is
the well known Traveling Salesman Problem (TSP) which is 
formally defined as follows: 

\vspace{0.1in}

\noindent {\bf TRAVELING SALESMAN PROBLEM } 

\noindent INSTANCE:  A finite set $C=\{c_1,c_2,\ldots,c_m\}$ of cities, 
and a distance $d(c_i,c_j)\in Z^+$ for each pair of cities $c_i,c_j\in C$.

\noindent QUESTION: What permutation 
\[
[c_{\pi(1)},c_{\pi(2)},\ldots,c_{\pi(m)}]
\]
of $C$ will minimize the tour length 
\[
\left\{
\sum_{i=1}^{m-1} d(c_{\pi(i)},c_{\pi(i+1)})\right\} + d(c_{\pi(m)},c_{\pi(1)})
\ ?
\]

Wille and Vennik \cite{wille85} used TSP to prove solving a homonuclear 
instance of DCP is NP-hard.  It is worthwhile to examine their proof in 
some detail since our proof of NP-hardness for the heteronuclear case follows 
a similar line of reasoning.

Instances of DCP can be expressed in graph-theoretical terms using a 
graph $G=(V,E)$ with vertex set $V$ and edge set $E$.  The graph $G$ 
is complete (i.e., the edge $e=\langle u,v\rangle\in E \; \forall \; u,v\in V$).  
Furthermore, $|V|\gg N$.  The edges are assigned weights $w(e) \; \forall \; e\in E$ 
where the weights reflect the interaction between vertices. An instance of 
DCP is therefore equivalent to selecting $V'\subset V$ with $|V'|=N$ so that

\begin{equation}
{\frac {1}{2}} \sum_{\begin{array}{c}\footnotesize u,v\in V' \\ 
e=\langle u,v\rangle\end{array}} w(e) = \mbox{minimal}
\end{equation}

\begin{figure}[htbp]
\centerline{\epsffile{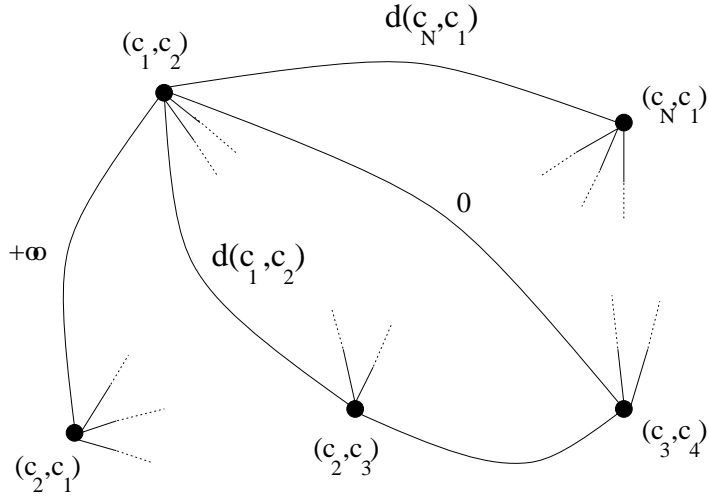}}
\caption{An example graph.  The graph is dense and edges are
weighted as given by Eq. (\ref{eqn2}).  The problem 
is to select $N$ vertices so
that the sum of the weighted edges is minimized.}
\label{tsp}
\end{figure}

The proof of Wille and Vennik uses an
undirected graph $G=(V,E)$ 
with $|V|=N(N-1)$ vertices (see Figure \ref{tsp}). 
Each vertex is labeled with ($c_i,c_j$) where $i,j=1,\ldots,N$ 
and $i\ne j$.  This indicates city $c_j$ is visited immediately after
city $c_i$.  Edges are then unordered pairs of the form 
\[
e=\langle(c_i,c_j)(c_k,c_l)\rangle
\]
with edge weights
\begin{equation}
w(e) \; = \; 
\left\{
\begin{array}{ll}
0 & \hspace{2pt}\mbox{if} \hspace{6pt} i\ne k, j\ne l \\
+\infty & \left\{ \begin{array}{l}
\hspace{2pt} \mbox{if} \hspace{6pt} i=k, j\ne l \\
\hspace{2pt} \mbox{if} \hspace{6pt} j=l, i\ne k \\
\hspace{2pt} \mbox{if} \hspace{6pt} l=i, k=j 
\end{array} 
\right. \\
d(c_k,c_l) & \hspace{2pt} \mbox{if} \hspace{6pt} l=i, k\ne j \\
d(c_i,c_j) & \hspace{2pt} \mbox{if} \hspace{6pt} k=j, l\ne i \\
\end{array}
\right.
\label{eqn2}
\end{equation}
Selecting $N$ vertices, 
$\{(c_{\pi(i)},c_{\pi(i+1)}); \; i=1,\ldots,N, c_{\pi(N+1)}=c_{\pi(1)}\}$,
so that the sum of the weights is minimal is equivalent to finding a
minimal length tour $\langle c_{\pi(1)},c_{\pi(2)},\ldots,c_{\pi(N)}\rangle$
thus solving an instance of TSP.  By restriction \cite{garey79}, this also
makes DCP NP-hard to solve.  This completes the Wille and Vennik proof.

The weight assignments given in Eq. (\ref{eqn2}) require some clarification.  
In TSP, weight is equivalent to distance whereas in DCP weight is
equivalent to pairwise potential energy.  Any plausible solution to an 
instance of TSP can visit each city but one time. 
In valid tour moves, the edge weight equals the intercity distance.
Disjointed tours have edges with zero weight---effectively removing that 
edge from the edge set $E$. This insures all tours can be defined as a
permutation of $m$ cities (see below).  An illegal tour move has an edge 
with infinite weight.  The total weight of the edges traversed in a tour 
measures the ``goodness'' of that tour; good tours have lower total edge 
weights.  More specifically,  

\begin{enumerate}
\item $i\ne k$ and $j\ne l$. This defines a disjointed tour where
the tour visits $c_i$ followed by $c_j$ and $c_k$ followed by
$c_l$.  However, it is not shown what other cities were visited
between $c_j$ and $c_k$.  Therefore, it is not possible to describe the
permutation and compute its tour length.     
\item $i=k$ and $j\ne l$. This defines an illegal tour that leaves the 
same city to visit two different cities.  
\item $j=l$ and $i\ne k$.  This defines an illegal tour where two distinct 
cities visit the same next city. 
\item $l=i$ and $k=j$.  This defines an illegal cyclic tour among only two
cities.  
\item $l=i, k\ne j$ or $k=j, l\ne i$.  These are legal tours.  
\end{enumerate}

\vspace{0.1in}

Each of the $N$ vertices in the minimal length tour is assigned a distinct atom 
from the cluster.  In homonuclear clusters \emph{all} one-to-one assignments of
atoms to these $N$ vertices are equivalent since the edge weights are based
only on interatomic distance.  This begins to explain why the Wille and Vennik 
proof \cite{wille85} and the Hendrickson algorithm \cite{hen95} cannot be extended 
to the heteronuclear case.  

First, consider a homonuclear cluster with $N>2$ atoms and suppose a new cluster 
is formed by swapping the spatial position of two atoms.  It is not possible to 
tell any physical difference between the two clusters because all atoms are 
identical and the interatomic distances remain unchanged.  Indeed, the new cluster 
will have a total energy identical to the original cluster because 
the pairwise interaction functions ($v(r_i-r_j)$ in Eq. (\ref{eqn1})) remain unchanged.  
Now suppose the cluster is composed of two distinct atom types, say Ar and Xe.  
The repulsive and attractive forces experienced by an Ar-Ar atom pair differs from 
those experienced by Ar-Xe or Xe-Xe pairs \emph{even if all pairs are separated 
by the same interatomic distance} \cite{pullan97}.  Swapping atom positions in 
heteronuclear clusters changes the type of atoms which interact, altering the 
individual interaction functions, and giving a different total energy to the new 
cluster.  

With homonuclear clusters it is acceptable to consider atoms as simple identical 
spheres where only interatomic distances contribute to the total energy.  It is
natural to model this system as an undirected graph where the edge weights reflect
forces derived solely from the interatomic distances. The search algorithm from 
\cite{hen95} takes this approach
thereby permitting the cluster to be optimized in stages by optimizing the relative
positions in subgraphs.  The complexity proof given in \cite{wille85} also made that 
assumption.  In fact, the graph used for that proof was constructed specifically without 
requiring any pair type information to set the edge weights.  That restriction was 
necessary to establish an equivalence between TSP and DCP.  

In heteronuclear systems, both atom type and distance determine pairwise forces so
the corresponding graph must have edge weights that take both distance and atom type 
into consideration.  
Even the relative positions of vertices from a subgraph cannot be optimized without the 
weights being set in this manner.  Consequently, search algorithms such as 
\cite{hen95} cannot be used for heteronuclear clusters because the overall cluster structure 
depends on connecting rigid bodies, formed from optimized subgraphs, where the edge weights
only consider interatomic distance.

To apply the proof of Wille Vennik \cite{wille85} to heteronuclear clusters, the
graph would have to be augmented with additional vertices.  For example, suppose
in the original (homonuclear) graph two vertices $i$ and $j$ are connected by
an edge $e=\langle i,j\rangle$. Then $w(e)$ is based solely on the interatomic distance 
between atoms $i$ and $j$.  The augmented graph would have a new vertex $j'$ where
the added edge $e'=\langle i,j'\rangle$ has a weight $w(e')$ computed from the interaction
of two dissimilar atoms, spaced at a distance $d(i,j')=d(i,j)$.
However, now the mere selection of 
$N$ vertices---sufficient to solve a homonuclear DCP---does not guarantee the correct 
mixture of atom types present in the heteronuclear cluster.  This restricts the proof from
\cite{wille85} to only homonuclear systems.

\section{Complexity in Heteronuclear Clusters} \label{4}

A different proof of complexity is needed for the heteronuclear clusters.  This 
new proof makes use of the following NP-hard problem \cite{garey79}:

\vspace{0.1in}

\noindent {\bf TRAVELING SALESMAN EXTENSION (TSE)} 

\noindent INSTANCE:  A finite set $C=\{c_1,c_2,\ldots,c_m\}$ of cities, a
distance $d(c_i,c_j)\in Z^+$ for each pair of cities $c_i,c_j\in C$, a
bound $B\in Z^+$, and a ``partial'' tour 
\[
\Theta = \langle c_{\pi(1)},c_{\pi(2)},\ldots,c_{\pi(K)}\rangle
\]
of $K$ distinct cities from $C$, $1\le K \le m$. 

\noindent QUESTION:  Can $\Theta$ be extended to a full tour
\[
\langle c_{\pi(1)},c_{\pi(2)},\ldots,c_{\pi(K)},c_{\pi(K+1)},\ldots,c_{\pi(m)}\rangle
\]
having total length $B$ or less?

\vspace{0.2in}

Consider a heteronuclear cluster which has a single atom of one type (denoted by 
$\alpha$) and $N-1$ atoms of a different type (denoted by $\beta$). A 
completely connected graph is $G=(V,E)$ with $|V|=N(N-1)$ is constructed with each vertex
labeled as described in Section \ref{3}.
Without loss in generality preassign an $\alpha$-atom to 
all vertices with labels ($c_1,c_j$), $j=2,\ldots,N$ and preassign
a $\beta$-atom to all remaining vertices.  

The search begins with a scan of all edges that touch
vertices with $\alpha$-atom assignments.  
Select the minimum weight edge.\footnote{As before, edges with weight 0
are effectively removed from the graph.}
This edge defines a partial tour $\langle c_{\pi(1)},c_{\pi(2)}\rangle$.
Now select $N-2$ more vertices, 
$\{(c_{\pi(i)},c_{\pi(i+1)}); \; i=3,\ldots,N, c_{\pi(N+1)}=c_{\pi(1)}\}$,
so that the sum of these weights is minimal.  (This search is limited to vertices
with $\beta$-atom assignments in order to maintain the proper mixture of atom types in
the corresponding cluster.)  Finding a minimal length full tour 
is thus equivalent to solving an instance of {\bf TSE}.  This proves that {\bf DCP} 
is NP-hard to solve for heteronuclear clusters as well.

\section{Discussion} \label{5}

The complexity in solving DCP forces researchers to use heuristic search 
algorithms.  Hill-climbing algorithms are not expected to do well because 
the PES has many local optima and it is highly likely the algorithm will 
quickly stop at one of them.  Conversely, stochastic search algorithms 
can be quite effective in such multimodal hypersurfaces.  It is therefore 
natural to ask if any one particular algorithm stands out as doing 
especially well against DCP. 

This question is not easy to answer and we need to turn to the No Free 
Lunch (NFL) Theorem \cite{wolpert97} for an explanation.  Essentially, 
this theorem says if some optimization (search) algorithm performs 
particularly well over a certain class of optimization problems, then 
it most likely will not perform as well over all remaining optimization 
problems.  This means one cannot choose, for example, simulated annealing 
to use for DCP just because it happens to work well for, say, scheduling problems.    
Direct comparisons between algorithms are insightful. But, without a 
conscientious attempt to make the comparisons fair, the results may be 
inconclusive, or in the worst case, be completely wrong \cite{gree97}.  
Monte Carlo techniques have been dominant in the area of cluster studies,
which is why newly proposed search algorithms are normally compared against 
them.

Most notable among the new algorithms are the evolutionary algorithms, which 
conduct searches that mimic Darwinian evolution: a ``population'' of clusters 
evolve to a low energy state by altering existing configurations via stochastic 
reproduction operators.  Natural selection determines which configurations survive 
to undergo further reproduction operations. 

Comparisons between evolutionary 
algorithms and Monte Carlo techniques have favored the former, although such 
comparisons sometimes lack sufficient mathematical rigor.  For example, Zeiri 
\cite{zeiri95} concluded that a genetic algorithm converges 
faster than simulated annealing after comparing the average of five runs---an 
unusually small sample size.  Normally results should be averaged over a 
considerably higher number of runs to help remove any potential bias in the 
random number generators used in the algorithms.  Nevertheless, there is growing
empirical evidence that says evolutionary algorithms consistently outperform the 
Monte Carlo techniques when applied against DCP \cite{hartke93}\cite{pullan97}
\cite{greexx}-\cite{deaven96}.  It appears as though evolutionary algorithms 
would be a good first choice search algorithm for cluster studies.

\section*{Acknowledgement}

The author wishes to thank the anonymous reviewers who made
several pertinent and valuable suggestions.

\end{document}